\documentclass[a4paper,11pt]{article}

\usepackage{jcappub} 
\usepackage{url}

\newcommand{\zaa}{{\em Astron.~Astrophys.}}

\newcommand{\zapj}{{\em Astrophys.~J.}}
\newcommand{\zapjl}{{\em Astrophys.~J.~Lett.}}
\newcommand{\zapjs}{{\em Astrophys.~J.~S.}}

\newcommand{\znat}{{\em Nature}}

\newcommand{\znp}{{\em Nucl.~Phys.}}
\newcommand{\zpl}{{\em Phys.~Lett.}}
\newcommand{\zpr}{{\em Phys.~Rev.}}
\newcommand{\zprd}{{\em Phys.~Rev. D}}
\newcommand{\zprc}{{\em Phys.~Rev. C}}
\newcommand{\zprl}{{\em Phys.~Rev.~Lett.}}

\newcommand{\zmnras}{{\em Mon. Not. R. Astron. Soc.}}

\newcommand{\zjcap}{{\em J. Cosmology Astropart. Phys.}}

\newcommand{\ob}{$\Omega_{\mathrm{b}}$}
\newcommand{\obh}{$\Omega_{\mathrm{b}}{h^2}$}
\newcommand{\deu}{D}
\newcommand{\tro}{$^3$He}
\newcommand{\qua}{$^4$He}
\newcommand{\six}{$^{6}$Li}
\newcommand{\sep}{$^{7}$Li}

\newcommand{\neu}{$^{9}$Be}
\newcommand{\dix}{$^{10}$B}
\newcommand{\onz}{$^{11}$B}
\newcommand{\hli}{$^4$He, D, $^3$He and $^{7}$Li}
\newcommand{\sbbn}{standard big bang nucleosynthesis}

\newcommand{\lap}{\mathrel{ \rlap{\raise.5ex\hbox{$<$}}
	            {\lower.5ex\hbox{$\sim$}}  } }

\title{\boldmath Standard big bang nucleosynthesis and primordial CNO Abundances after {\it{Planck}}}

\author[a,1]{Alain Coc \note{Corresponding author.}}
\author[b,c]{, Jean-Philippe Uzan}
\author[b,c]{and Elisabeth Vangioni}

\affiliation[a]{Centre de Sciences Nucl\'eaires et de Sciences de la
Mati\`ere  (CSNSM), CNRS/IN2P3, \\ Universit\'e~Paris~Sud~11, UMR~8609,
B\^atiment 104, F--91405 Orsay Campus (France)}
\affiliation[b]{Institut d'Astrophysique de Paris,
              UMR-7095 du CNRS, Universit\'e Pierre et Marie
              Curie,
              98 bis bd Arago, 75014 Paris (France)}
\affiliation[c]{Sorbonne Universit\'es, Institut Lagrange de Paris, 98 bis bd Arago, 75014 Paris (France).}

\emailAdd{coc@csnsm.in2p3.fr}
\emailAdd{uzan@iap.fr}
\emailAdd{vangioni@iap.fr}

\abstract{
Primordial or big bang nucleosynthesis (BBN)  is one of the three historical strong evidences for the big bang model. The recent results by the {\it{Planck}} satellite mission have slightly changed the estimate of the  baryonic density compared to the  previous WMAP analysis. This article updates the BBN predictions for the light elements using the cosmological parameters determined by {\it{Planck}}, as well as an improvement of the nuclear network and new spectroscopic observations. There is a slight lowering of the primordial Li/H abundance, however, this lithium value still remains typically 3 times larger  than its observed spectroscopic abundance in halo stars of the Galaxy. 
According to the importance of this "lithium problem", we trace the small changes in its BBN calculated abundance following updates of the baryonic density, neutron lifetime
and networks.  
In addition, for the first time, we provide confidence limits for the production of \six, \neu, \onz\ and CNO, resulting from our extensive Monte Carlo calculation with our
extended network.
A specific  focus is cast on CNO primordial production. Considering uncertainties on the nuclear rates around the CNO formation, we obtain $\rm{CNO/H} \approx (5-30)\times10^{-15}$. 
We further improve this estimate by analyzing correlations between yields and reaction rates and identified new influential 
reaction rates. These uncertain rates, if {\em simultaneously} varied could lead 
to a significant increase of CNO production:  $\rm{CNO/H}\sim10^{-13}$. 
This result is important for the  study of population III star formation during the dark ages.
}

\begin{document}
\maketitle
\flushbottom

\section{Introduction}

There are three historical observational evidences for the big bang model: the cosmic expansion, the Cosmic Microwave Background (CMB) radiation and primordial or big bang nucleosynthesis (BBN). Today, they are complemented by a large number of evidences in particular from the properties of the large scale structures (see e.g. \citet{pubook} for a textbook description). BBN predicts the primordial abundances of the ``light cosmological nuclei'': \hli\ that are produced during the first 20 min after the big bang when the Universe was dense and hot enough for nuclear reactions to take place (see e.g. \citet{Ste07,Ioc09,fields11} for recent reviews). 
The comparison of the calculated and observed abundances shows an overall good agreement except for the \sep. The essential cosmological parameter of the model is the baryonic density $\Omega_{\rm b}$. It is related to the baryon to photon ratio, 
$\eta\equiv n_{\rm b}/n_{\gamma}=2.738\times10^{-8}\;\Omega_{\mathrm{b}}h^2$ (see the appendix) that
remains constant during the expansion after the electron--positron annihilation.
\obh\ is now well measured  from the angular power spectrum of the CMB temperature anisotropies. A precise value for this, previously free, parameter  was provided by the Wilkinson Microwave Anisotropy Probe (WMAP9) satellite, $\Omega_{\rm b}h^2=0.02243\pm0.00055$,~
("Nine-year (MASTER)", \citet{WMAP9}) while the recent {\it{Planck}} mission updated it to $\Omega_{\rm b}h^2$=0.02218$\pm$0.00026 ("{\it{Planck}}+lensing+WP+highL", 
\citet{Planck13}). This value is chosen because it includes all the last cosmological constraints. We calculate here the \hli\ primordial abundances by Monte Carlo, using our extended 424 nuclear reaction network \citep{Coc12a}, also taking into account the updated value of the neutron lifetime \cite{PDG12}.
In \sbbn, only traces of other isotopes are produced: \six, \neu, \dix, \onz\ and CNO.   
The CNO abundance is of peculiar interest since it may affect Pop III stellar evolution in the first structures of the Universe.  The value which could impact this evolution is estimated to be 10$^{-11}$ \cite{Cas93}  or
even as low as  10$^{-13}$ (in number of atoms relative to hydrogen, CNO/H) for the less massive stars \cite{Eks08}. In this context, it is important to evaluate carefully the  BBN CNO abundance.
In our previous work \citep{Coc12a} we obtained a much lower value CNO/H=$0.7\times10^{-15}$ but no upper nor lower limit (see also Ref. \cite{Ioc07}).
In this paper, we use the results of our Monte Carlo calculations $i$) to estimate the uncertainties on the BBN production of the minor isotopes,
and in particular of CNO  and $ii$) analyze the correlations between reaction rates and isotopic abundances to identify potentially important
reactions that were not identified in our previous sensitivity analysis.
We show that by calculating correlations, we find important reactions that were overlooked in sensitivity studies
changing one reaction at a time. This is crucial because the level of the CNO abundance plays a key role in the evolution of the first stars.

\section{Primitive observational  abundances: update}

\subsection{\hli\ observations}
\label{s:obsheli}

Deuterium is a very fragile isotope, easily destroyed after BBN. Its most primitive abundance is determined from the observation of cosmological clouds at high redshift, on the line of sight of distant quasars. Very few such  observations  are available  \cite{pettini08}. 
Up to now, the observation of about 10  quasar absorption systems  gave the weighted mean abundance of deuterium D/H = $(3.02 \pm 0.23) \times 10^{-5}$~\citep{olive2012}.  However, these individual measurements of D/H show a considerable scatter and it is likely that systematic errors dominate the uncertainties. 
Recently, Cooke et al.~\citep{pettini2012, cooke2014} have done new observations of Damped Lyman-$\alpha$ (DLA) systems at high redshift 
and made a global reanalysis, including previous observations, that lead to  a new mean value of 
\begin{equation}
{\rm D/H} = (2.53 \pm 0.04) \times 10^{-5},
 \end{equation}
lower and with a much narrower error bar than in previous determinations.

After BBN, \qua\ is still produced by stars. Its primitive abundance is deduced from observations
in H{\sc ii} (ionized hydrogen) regions of compact blue galaxies. The primordial \qua\ mass fraction, $Y_p$, is obtained from the extrapolation to zero metallicity but is affected by systematic uncertainties \citep{aver10, isot10} such as plasma temperature or stellar absorption. 
Recently, \citet{aver12, aver13},  using a subset of the data set found in  Izotov et al (\cite{isot07,izotov13}, and references therein) have incorporated new atomic data and updated their recent Markov Chain Monte Carlo analysis; so they have determined the primordial helium abundance by a regression to zero metallicity (however within a narrow range of metallicity),  
\begin{equation}
Y_p = 0.2465 \pm 0.0097
\end{equation}
which corresponds to a narrower error bar than previous constraints.  This is the value we use to compare with our calculations.
Another recent determination of \citet{izotov13}, $Y_p = 0.254 \pm 0.003$ is higher than the \citet{aver13} value.
The difference comes from a different a regression to zero metallicity (the mean value, i.e. with no regression, $Y_p = 0.2535 \pm 0.0036$ of \citet{aver13} and 
the \citet{izotov13} one are in perfect agreement) and differences in the atomic physics involved. 

Contrary to \qua\ ,  \tro\ is both produced and destroyed in stars throughout its galactic evolution, so that the evolution of its
abundance as a function of time is subject to large uncertainties.
\tro\ has been observed in our Galaxy \citep{Ban02}, and one only gets a 'local' constraint
\begin{equation}
\hbox{\tro/H}  = (1.1 \pm 0.2) \times 10^{-5}.
\end{equation}
This observation is consistent with the BBN predicted value, nevertheless, due to these uncertainties related to stellar evolution, it is difficult to use it as a constraint.

Primitive lithium abundance is deduced from observations of low metallicity stars in the halo of our Galaxy where the lithium abundance is almost independent of metallicity, displaying the so-called Spite plateau \citep{Spite82}. This interpretation assumes that lithium has not been depleted at the surface of these stars, so that the presently observed abundance can be assumed to be equal to the primitive one. The small scatter of values around the Spite plateau is indeed an indication that depletion may not have been very efficient. However, at very low metallicity, on top of a lot of scatter, a slight decrease of Li with metallicity appears.  In this context, \citet{Spite10} pointed out  that the abundance of lithium could be even lower when extrapolating toward zero-metal stars, just after the Big Bang. So, considering the Spite plateau, there is a discrepancy between the value {\it{i)}} deduced from these  observed spectroscopic abundances and {\it{ii)}} the  BBN theoretical predictions assuming  \ob\
  is determined by the CMB observations.  Many studies have been devoted to the resolution of this so-called {\it Lithium problem} and many possible ``solutions'', none fully satisfactory, have been proposed. For a detailed analysis see \citet{fields11}, the proceedings of the meeting ``Lithium in the cosmos''~\citep{LiinC} and recently \citet{cyburt13}.
Note that recent lithium observations \cite{How12} have been done in the Small Magellanic Cloud which is a nearby irregular galaxy with quarter of the sun's metallicity and
its abundance is found to be nearly equal to the BBN one. It could be a strong constraint for the lithium galactic evolution.
Astronomical observations of these metal poor halo stars \citep{Ryanetal2000} have thus led to a relative primordial abundance of
${\rm Li/H}= (1.23^{+0.34}_{-0.16}) \times 10^{-10}$ while a more recent analysis~\citep{sbordone10} gives
\begin{equation}
  {\rm Li/H}=  (1.58 ^{+0.35}_{-0.28}) \times 10^{-10}
\end{equation}  
which we use in our analysis. For reviews on the  Li observations, we refer to~\citet{Spite10} and \citet{frebel13}. 

\subsection{\six, \neu, B, CNO observations} 
\label{s:obslibeb}

The origin of the light elements LiBeB, is a crossing point between optical and gamma spectroscopy, 
non thermal nucleosynthesis (via spallation with Galactic Cosmic Rays, GCR), stellar evolution and Big Bang nucleosynthesis.
Lithium-6  is also observed in metal poor stars, as discussed above for \sep. 
Its observational history is peculiar. 
In the past, \citet{asplund06} have provided observations of  \six/\sep\  ratio
 which suggesting the presence of a plateau, at typically   \six/H = $ 10^{-11}$,
 leading to a possible pre-galactic origin of this isotope (see  \citet{rollinde05}). 
  In a second time,
  the observational \six\  plateau has been questioned in  \citet{cayrel07} 
  who have considered in more details the line shape asymmetries. 
  Consequently, there has been a retraction concerning the existence of a  \six/H  plateau \cite{ lind13}.
  Presently, only one star, HD84937, presents a  \six/\sep\  ratio of the order of 0.05
  (see  \citet{steffen12}) and there is no remaining evidence for a plateau at very low metallicity.
  We hence use this value as an upper limit: \six/H$\lap10^{-11}$  while the prediction of the BBN calculations are \six/H =  $10^{-14}$.

Beryllium has only one stable isotope \neu. As D and Li, it is a fragile nucleus. It is formed in the vicinity of Type II supernovae (SNII) by non thermal process (spallation) (see \citet{vangioni98, vangioni00}). It is also observed in metal poor stars. \citet{boesgaard11} (and references therein) have performed an update of Be observations in metal poor stars which provides a primitive abundance at very low metallicity of the order of  $ {\rm Be/H}=  3.\times 10^{-14}$
at   [Fe/H] = -3.5. This observation,  that we adopt as upper limit of the primordial abundance, has to be compared to the typical primordial Be abundance,  $ {\rm Be/H}=  10^{-18}$.

Boron has two stable isotopes:  \dix\   and \onz. It is also synthesized by non thermal processes, GCR or neutrinos (for  \onz) (see \citet{vangioni96, vangioni98, vangioni01}). The most recent observations of boron in low  metal stars come from 
\citet{duncan97} and  \citet{garcialopez98}. 
In the galactic halo, the lowest boron abundance at  [Fe/H] $\lap$ -3. is  
B/H$\approx10^{-12}$, to be compared to the typical primordial B abundance $ {\rm B/H}=  3.  \times 10^{-16}$.

For a general review of these light elements, see the IAU Proceedings \citet{charbonnel10}.

Finally, CNO elements are observed in the lowest metal poor stars (around [Fe/H]=-5). The observed abundance of CNO is typically [CNO/H]= -4, 
relatively to the solar abundance i.e. primitive CNO/H$<10^{-7}$. For a review see \citet{frebel13} and references therein.

\section{Effects of \obh, $\tau_{\rm{n}}$ changes and network extension}
\label{s:tables}

Concerning the update of the CMB, a comparison between the columns of
Table 1 shows the effect of a change in \obh\ from \citet{Spe07}  (column 2) to \citet{Kom11} (columns 3 or 4; depending on the choice of the neutron
lifetime) and to  \citet{Planck13} (column 5).
In that way, we can trace the changes in our previous publications e.g. in \citet{CV10} where we used three-years WMAP only \citep{Spe07}, 
in \citet{Coc12a} (seven-years WMAP  \citep{Kom11}) until this work ({\it{Planck}}+lensing+WP+highL \citep{Planck13}), together with
evolving Particle Data Group evaluation of the neutron lifetime, over the years.
For the final calculations, we choose, for \obh, the value from the
Planck paper that incorporate the Planck temperature data, polarization
data from the WMAP satellite in the multipole range $2 < \ell < 23$, 
information from the lensing potential as determined from the trispectrum
computed on Planck's maps \citep{Planck13xvii} and
information
coming from ground-based high resolution experiments, such as ACT and SPT.
This data set, referred to as ``Planck+lensing+WP+highL"  can be considered
to the most uptodate combination of CMB data, hence leading to the most
accurate estimation of the cosmological parameters.
These changes mostly affects \sep/H by about 4\% and D/H by about 2.7\% while the other changes are below a percent.
Even though it won't change the nature of the  "lithium problem", we found important to  trace the small changes in its BBN calculated abundance 
following updates of the baryonic density, neutron lifetime and networks.  
A BBN evaluation has been done by \citet{Planck13}, using \obh = 0.02207{$\pm$0.00027};  their prediction regarding the $Y_p$  and \deu/H abundances are similar to ours (0.24725$\pm$0.00032 and   2.656$\pm$0.067 $ \times10^{-5}$ respectively at $\eta_\mathrm{CMB}$) but they do not provide any \sep/H value. 
In Table~\ref{t:small}, we show the influence of changes in \obh\ and $\tau_{\rm{n}}$ with a minimal network. Table~\ref{t:large}, compared to Table~\ref{t:small},
show the effect of extending the network. Small differences in D and \sep\ are observed but that could not be traced to well identified origins:
as we shall see in \S~\ref{s:cno}, there are combined effects of reaction rates that are different from the effects of individual rates. 
However, changes in neutron late time abundance and sub--dominant $^7$Be destruction mechanism like $^7$Be(n,$\alpha)^4$He \citep{Wag69} play a role. 
Finally, in Table~\ref{t:obs}, we compare our Monte Carlo (\S~\ref{s:mc}) results
to observations (\S~\ref{s:obsheli}).

\begin{table*}[htbp!] 
\caption{\label{t:hli} Primordial abundances with reduced network. (Bold face displayed values highlight parameter changes.)}
\begin{center}
\begin{tabular}{ccccccc}
\hline
  &  (a)  &  (b)   & This work  & This work \\
\hline  
Nb. reactions & 13  & 15 & 15 & 15 \\
\hline
 \obh & $0.0223^{+0.00075}_{-0.00073}$ (c) & {\bf 0.02249} (e) & 0.02249 (e) & {\bf 0.02207 (g)}  \\
 $\tau_{\rm{n}}$ & 885.7$\pm$0.8 (d) & 885.7 (d) & {\bf 880.1 (f)} &  880.1 \\
 \hline
$Y_p$*     &  0.2476$\pm$0.0004       & 0.2475    & 0.2464 & 0.2462 \\
 \deu/H   ($ \times10^{-5})$& $2.68\pm0.15$    & 2.64   &   2.64 & 2.72 \\
 \tro/H    ($ \times10^{-5}$) & 1.05$\pm$0.04    & 1.05      &   1.05 & 1.06 \\
 \sep/H ($\times10^{-10}$)  &  5.14$\pm$0.50 &  5.20  & 5.18 & 4.98 \\
\hline
\end{tabular}\\
\; *An additional ${\Delta}Y_p=0.0018$ correction should be made (see text).
(a) \citet{CV10}; (b) \citet{Coc12a},
(c) \citet{Spe07} ; (d) \citet{PDG08};\\ (e) \citet{Kom11} ; (f) \citet{PDG12}; (g) \citet{Planck13}
\end{center}
\label{t:small}
\end{table*}

\begin{table*}[htbp!] 
\caption{\label{t:hlix} Primordial abundances with extended network.}
\begin{center}
\begin{tabular}{cccccc}
\hline
 Nb. reactions  & 424 & 424 & 424  \\\hline
\obh  & 0.02243 (x) & 0.02207 (y)  & 0.02218$\pm$0.00026 (z)\\
\hline
 $\tau_{\rm{n}}$ & 880.1& 880.1& 880.1$\pm$1.1 \\
\hline
$Y_p$*      &0.2465 &0.2463& 0.2461--0.2466  \\
\deu/H   ($ \times10^{-5})$&2.60 &2.67 & 2.57--2.72\\
\tro/H    ($ \times10^{-5}$)  & 1.04  &1.05 & 1.02--1.08 \\
\sep/H ($\times10^{-10}$)   &5.13  &4.96&   4.56--5.34 \\
\hline
\end{tabular}\\
\; *An additional ${\Delta}Y_p=0.0018$ correction should be made (see text). Hinshaw et al., WMAP9  \citep{WMAP9} 
(y) {\it{Planck}} only \citep{Planck13} 
(z) {\it{Planck}}+lensing+WP+highL \citep{Planck13} 
\end{center}
\label{t:large}
\end{table*}

Note that in order to precisely compare our $Y_p$ values with \citet{Planck13} and some other works, 
an additional ${\Delta}Y_p=0.0018$ correction should be made. The weak reaction rates \citep{Dic82}
that we numerically integrate include zero-temperature Coulomb and radiative corrections 
\citep{Dic82,Lop99}. This correction amounts to  ${\Delta}Y_p=0.00316$, in our calculations at  the
relevant \obh\ (0.0031 in \citet{Lop99}). 
What we have neglected, up to now are the finite-nucleon mass correction (${\Delta}Y_p=0.0012$ \citep{Lop99}), 
finite-temperature radiative correction (${\Delta}Y_p=0.0003$ \citep{Lop99}), QED plasma (${\Delta}Y_p=0.0001$ 
\citep{Lop99}) and neutrino decoupling (${\Delta}Y_p=0.0002$ \citep{Man05}), for a total of ${\Delta}Y_p=0.0018$,
because they cannot be easily directly re-calculated\footnote{
We numerically calculate  the weak rates, to keep track on the dependence w.r.t. $G_F$ (Fermi constant), $m_e$ (electron mass) and $Q_{\mathrm np}$
(neutron-pronton mass difference), essential in our investigations concerning variations of constants in BBN \citep{cstes07,cstes12}.  
}.   
Hence, our quoted $Y_p$ uncertainties reflect the nuclear uncertainties, mainly $\tau_{\rm{n}}$, not the theoretical uncertainties 
on theses corrections, difficult to estimate for us.  Nevertheless, these neglected corrections  (i.e. the ${\Delta}Y_p=0.0018$ not included in the above Tables) remain
one order of magnitude below the observational uncertainty:  $ 0.2465 \pm 0.0097$.

Since the neutron lifetime and baryonic density values are both subject to debate (\citet{PDG12,Wie11}  for $\tau_\mathrm{n}$ and 
\citet{Planck13} and \citet{Spe13} for \obh), instead of providing new tabulated values, we propose fits for the BBN abundances of \hli\ abundances 
as a function of \obh, ${\tau_{\rm{n}}}$ and ${N_{{\rm{eff}}}}$ (see definition below) hence of the BBN predictions.
This can be used to update any column of Tables~\ref{t:hli} and \ref{t:hlix} with  \obh\ or ${\tau_{\rm{n}}}$ different values than those in the same column
or for $\Delta{N_{{\rm{eff}}}} = {N_{{\rm{eff}}}} -3\neq0$. 
Our motivation is to provide simple fits that could be directly used to calculate
the effect induced by the small changes in  and  \obh\ or ${\tau_{\rm{n}}}$  until their precise values are settled.   
Fitted BBN abundances were already provided by  e.g \citet{Lop99,Ioc09,Ste12} but, they $\eta$ instead of  
\obh\ \citep{Lop99} (see the appendix), lack $N_{{\rm{eff}}}$ \citep{Ioc09} or ${\tau_{\rm{n}}}$ \citep{Ste12}.
We checked that, for small variations our results are very close to \citet{Ioc09} for ${\tau_{\rm{n}}}$  and  \obh\
dependence\footnote{After a typo, 0.39$\to$0.039 in  \citet{Ioc09}, Eq. (63), has been corrected.}.

\begin{eqnarray}\nonumber
\Delta Y_p &= &+0.4274\;\Delta\Omega_{\mathrm{b}}h^2+2.043\times10^{-4}\;\Delta\tau_{\rm{n}}+\\
&&1.348\times10^{-2}\;\Delta{N_{{\rm{eff}}}}-9.805\times10^{-4}\;\Delta{N^2_{{\rm{eff}}}} 
 \end{eqnarray}
 
\begin{equation}
\Delta\mathrm{D/H}  = -1.878\times10^{-3}\;\Delta\Omega_{\mathrm{b}}h^2+1.256\times10^{-8}\;\Delta\tau_{\rm{n}}+3.564\times10^{-6}\;\Delta{N_{{\rm{eff}}}},
\end{equation}

\begin{equation}
\Delta\mathrm{^3He/H} = -2.783\times10^{-4}\;\Delta\Omega_{\mathrm{b}}h^2+1.761\times10^{-9}\;\Delta\tau_{\rm{n}}+ 5.064\times10^{-7}\;\Delta{N_{{\rm{eff}}}},
\end{equation}

\begin{equation}
\Delta\mathrm{^7Li/H} = +4.767\times10^{-8}\;\Delta\Omega_{\mathrm{b}}h^2+2.4541\times10^{-13}\;\Delta\tau_{\rm{n}}-4.686\times10^{-11}\;\Delta{N_{{\rm{eff}}}}.
\end{equation}

These fits also consider the variations induced by a change in $-1<\Delta{N_{{\rm{eff}}}}<+1$ where $N_{{\rm{eff}}}$, the effective number of relativistic
degrees of freedom,  is defined by:
\begin{equation}
 \rho_{\rm r}= \left[1+\frac{7}{8}N_{\rm eff}\left(\frac{T_\nu}{T}\right)^4 + \frac{15}{\pi^2}\frac{\rho_e}{T^4}
 \right]\rho_\gamma.
 \label{q:neff}
\end{equation}
At recombination, ${\rho_e}\ll\rho_{\rm r}$ and $\left({T_\nu}/T\right)^4=\left(4/11\right)^{4/3}$, so that Eq.~(\ref{q:neff}) matches the definition used
in CMB analyses.

\section{Method}

\subsection{Monte Carlo}
\label{s:mc}

\citet{CV10} used a network reduced to the 12 main reactions (13 with the $^3$H(p,$\gamma)^4$He that plays a negligible role) for which the rate uncertainties are small compared to all other ones, and sampled the rates within the uncertainty range according to a normal distribution. 
Here, the extended network includes reaction rates that can be uncertain by a factor of a few orders of magnitude due to the lack 
of experimental data. Hence, we follow \citet{Eval1} and use a lognormal distribution  to cope with these large uncertainty factors 
together with ensuring that the sampled rates are positive: 
\begin{equation}
f(x) = \frac{1}{\sigma \sqrt{2\pi}} \frac{1}{x} e^{-(\ln x - \mu)^2/(2\sigma^2)}  \label{lognormalpdf}
\end{equation}
(with $x\equiv{N_A}\langle\sigma{v}\rangle$ for short).
This is equivalent to assumption that  $\ln(x)$ is Gaussian distributed with expectation value $\mu$ and variance $\sigma^2$. For the lognormal distribution, one has:  
\begin{equation}
E[x] = e^{(2\mu + \sigma^2)/2},~~\mathrm{and}~~
V[x] = e^{(2\mu + \sigma^2)}~\left[e^{\sigma^2} - 1 \right]. \label{logon}
\end{equation}
As discussed in \citet{Eval1} (see their Fig.~1), for small $\sigma$ a lognormal distribution and a normal distribution with the same 
expectation value and variance are close to each other.
Hence, since the uncertainty in the 12 main reaction rates are small, using here a lognormal distribution for those reactions makes no difference 
with \citet{CV10} results.

To perform the Monte Carlo calculation, we follow the prescription of \citet{Sal13}. 
Namely the reaction rates $x_k$ $\equiv{N_A}\langle\sigma{v}\rangle_k$, (with $k$ being the index of the reaction), 
are assumed to follow a lognormal distribution:
\begin{equation}
x_k(T)=\exp\left(\mu_k(T)+p_k\sigma_k(T)\right)
\label{q:ln}
\end{equation}
where  $p_k$ is sampled according to a {\em normal} distribution of mean 0 and variance 1 (Eq.~(22)  of \citet{Sal13}).
$\mu_k$ and $\sigma_k$ determine the location of the distribution and its width which are tabulated as a function of $T$:
\begin{equation}
x_{\rm med}\equiv\exp\left(\mu\right) 
\label{q:xmed}
\end{equation}
is the median rate and 
\begin{equation}
f\equiv\exp\left(\sigma\right)
\label{q:fu}
\end{equation}
the uncertainty factor. 
They are deduced from the evaluation of rate uncertainties. 
For reactions for which ``high" and ``low" rates\footnote{In the literature one often find tables of reaction rates with labels such as ``high", ``low",
``upper", ``lower" or "recommended". In most recent works e.g. \citet{Des04,Eval2}, they have well defined statistical significance. In many older works,
e.g. \citet{NACRE} they have no precise definition but we still use Eqs.~(\ref{q:mu}) and (\ref{q:sigma}), for lack of anything better, if this is the only source,
to calculate $\mu$ and $\sigma$, to be used in the Monte Carlo.
(See \citet{Eval1}.)}
are
available, 
\begin{equation}
\mu\equiv\ln\sqrt{x_{\rm{low}}\times x_{\rm{high}}}
\label{q:mu}
\end{equation}
and  
\begin{equation}
\sigma\equiv\ln\sqrt{x_{\rm{high}}/x_{\rm{low}}}
\label{q:sigma}
\end{equation}
(see \citet{Eval1}). 
To avoid erratic numerical behavior, we limit the sampling to values lower than one thousand times the median rate, i.e.,
\begin{equation}
x<10^3\;x_{\rm med}, 
\label{q:xmax}
\end{equation}
to remain within a range already explored \citep{Coc12a}.

The Monte Carlo calculation proceeds as follows.
For each trial labeled by $i$, we sample randomly a set of $\{p_{k;i}\}$ different numbers where $k$ that runs from 1 to $N$ (number of reactions) 
is the index of the reaction.
Each one follows independently a Gaussian distribution of mean value 0 and variance 1 and is obtained from a standard random number generator. 
A BBN calculation is performed with the set of reaction rates $\{x_{k;i}\}$ obtained from Eq.~(\ref{q:ln}) that produce the set of isotopic abundances  
$\{y_{j;i}\}$ obtained in trial $i$. Here, $j$=\qua, \deu, \tro, \sep, \six, \neu, \onz\ and CNO  is the index the corresponding abundances 
after decays of radioactive isotopes ($^7$Be, $^3$H, $^{11}$C, \dots)  or summation of A$\geq$12 isotopic abundances (CNO).
For further analyses, for each trial, not only the final abundances are stored in a database but also all the  $\{p_{k;i}\}$ values.
This allows for correlation studies discussed below.

\begin{figure}[htb!]
\begin{center}
 \includegraphics[width=.8\textwidth]{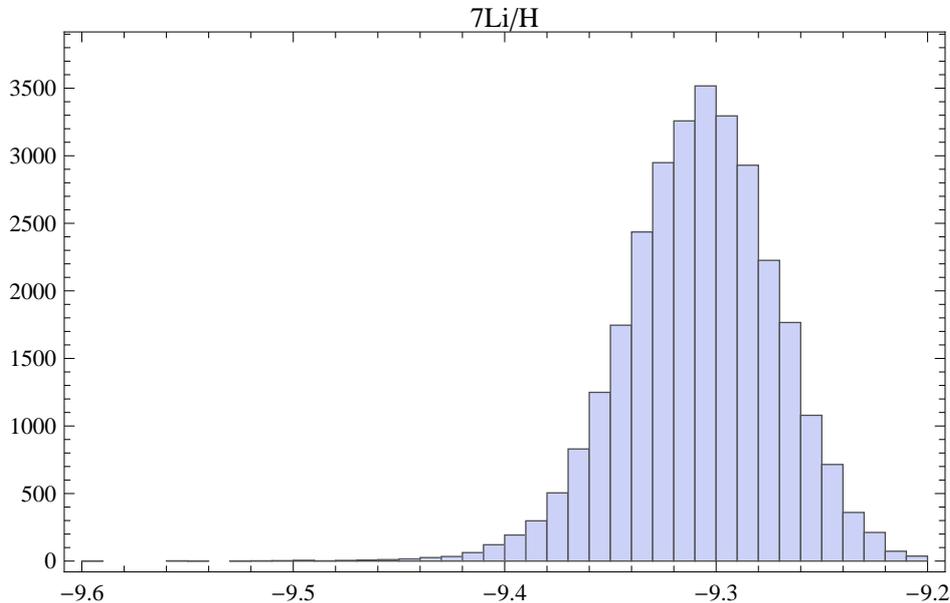}
\caption{\label{f:hsep} Histogram of the\sep/H distribution, from which the median and confidence interval is calculated.}
\end{center}
\end{figure}

To obtain the primordial abundances and their uncertainties, as tabulated below, \obh\ also is randomly sampled, 
following a Gaussian distribution according to the CMB deduced data.
After 30000 such computations, the calculated distributions of abundances are obtained as displayed in Fig.~\ref{f:hsep} for \sep. 
The median primordial abundances and associated 68\% confidence intervals are then calculated  
by taking respectively the 0.5,  0.16 and  0.84 quantiles of the abundance distributions (see Fig.~5 in \citet{Eval1}). 

It has been suspected that traditional sensitivity studies, in which only one reaction is varied while the others are held constant,
cannot properly address all the important correlations between rate uncertainties and nucleosynthetic predictions \cite{Stats}. 
Searching for such correlations was done by \citet{Par08} for X--ray bursts (see their Figs.~7 and 8). 
They have not found new influential reaction, by examining correlations, as compared to their more traditional sensitivity studies.
However, we think that it is worth applying this methods to BBN because the density
of the ``fabric" of a BBN network (d, t and \tro\ in addition to the usual n, p and $\alpha$--particles induced reactions) is higher, offering more potential paths.
For this purpose, the same Monte Carlo calculation is performed, except that \obh\ is fixed, to obtain a data base that can be used to study correlations. 
The Pearson's correlation coefficient between isotope $j$ and reaction $k$ is, then, calculated as:
\begin{equation}
C_{j,k}=\frac{1}{N}\frac{
\sum_{i=1}^{N}\left(y_{j;i}\;p_{k;i}-\overline{y_j}\;\overline{p_{k}}\right)}
{\sigma_j\sigma'_k}
\label{q:cor}
\end{equation}
where $y_{j;i}$ is again the final abundance of isotope $j$ obtained in trial $i$ with the set of reaction rates $\{x_{k;i}\}$ obtained from Eq.~(\ref{q:ln})
with the set of randomly sampled $\{p_{k;i}\}$.
The correlation coefficients are calculated, with  $y_j\equiv\ln(n_j/n_H)$ for $j\neq$\qua\ and $y_\mathrm{^4{He}} {\equiv}Y_p$ for the abundances. 
In Eq.~(\ref{q:cor}),  $\overline{y_j}$ [$\overline{p_k}$]  and $\sigma_j$ [$\sigma'_k$] stand for the mean and standard deviation of $y_j$ [$p_k$].  
If it were not for the condition (\ref{q:xmax}), one would have obviously  $\overline{p_k}\equiv0$ and $\sigma'_k\equiv1$ since the $p_k$ are sampled 
according to a normal distribution.

\subsection{Reaction rates and uncertainties}
\label{s:rates}

In this study, the reaction network and the thermonuclear rates comes from \citet{Coc12a}. 
Namely, it includes 59 nuclides from neutron to $^{23}$Na, linked
by 391 reactions involving n, p, d, t, \tro\ and $\alpha$--particles induced reactions and 33 $\beta$-decay processes \citep{Aud03}. 
Reaction rates
were taken primarily from \citet{NACRE,Des04,Eval2,NACRE2} and other evaluations
when available. 
Following their sensitivity study a few reaction rates were re-evaluated by \citet{Coc12a}; they are also used here.
The complete list of reactions with associated references to the origin of the rates can 
be found in Table~4 of  \citet{Coc12a} (except for $^7$Be(n,$\alpha)^4$He for which we use here the \citet{Wag69} rate instead of the TALYS one in \citet{Coc12a}). 
Since our previous Monte Carlo BBN calculations \citep{CV10}, no change has been made concerning 11 of the 12 main BBN reactions rates.  
We use those from the evaluation performed by  \citet{Des04} except for $^1$H(n,$\gamma$)D \citep{And06} and 
$^3$He($\alpha,\gamma)^7$Be \citep{Cyb08a}. A new experiment \citep{Leo06} have provided new data for the D(d,n)$^3$He and D(d,n)$^3$He    
cross section after the \citet{Des04} evaluation. Within the BBN energy range, the new data fall exactly on the \citet{Des04} R--matrix fit
(see Fig.~2 of \citet{NN2012}) that has not yet been updated with this new data. 

A recent paper by \citet{THM} provides new evaluation of \hli\ primordial abundances based on new nuclear data,
but comparison with this work is difficult. The new data is extracted by an {\em indirect} experimental method (the 
Trojan Horse Method) that requires theoretical input. We use instead the results of {\em direct measurements that 
provide the same data at, and even below, BBN energies} \cite{Des04}, not affected by screening \citep{screening}. 
Besides, nothing is said of the origin for rate of the $^3$He($\alpha,\gamma)^7$Be used in their
calculations, which is known to be essential for \sep\ prediction, and has been updated by  \citet{Cyb08a}.    

The only modification of the main rates concerns the weak reactions involved in n$\leftrightarrow$p equilibrium  whose rates \citep{Dic82} is 
determined from the standard theory of the weak interaction but needs to be normalized to the experimental neutron  lifetime.  
The latter has recently been revised by the Particle Data Group from  
885.7$\pm$0.8~s~\citep{PDG08}, used in \citet{CV10}, to 880.1$\pm$1.1~s \citep{PDG12}.
This significant change is due to the inclusion of the \citet{Ser05} experimental value, now comforted by 
new analyses (see \citet{Wie11,PDG12} for more details), that was previously left out of the averaging because of its inconsistency with other data. 
For this quantity, we use the latest value recommended by the Particle Data Group 
$\tau_\mathrm{n}=880.1\pm1.1$~s \cite{PDG12}, but are aware that this remains an open debate \citep{Wie11} but
that affects essentially, only \qua.

The calculation of  \six/H depends directly on the D$(\alpha,\gamma)^6$Li reaction rate that was plagued with by  
large uncertainty  \citep{NACRE},
 New measurement of the D$(\alpha,\gamma)$ \six\, which is the  main way 
 to produce primordial \six\, have been performed. 
 In the absence of direct measurements at BBN energy, \citet{Ham10} used the Coulomb breakup technique
 to extract the cross section.  Very recently, a direct measurement has been performed at LUNA \citep{LUNA}.
 The results of this very difficult experiment agree well with those from the indirect method  \citep{Ham10}.
 So, this confirmed low cross section comforts the prediction of the BBN calculations of a low primordial \six\ value,  \six/H =  $10^{-14}$.

For the remaining of the 391 reactions, we use tabulated $\mu$ and $\sigma$ \citep{Eval2}, tabulated limits \citep{NACRE,Des04,NACRE2,Coc12a}
together with Eqs.~(\ref{q:mu}) and (\ref{q:sigma}) and for the others, tabulated rates together with estimated uncertainty factors and Eqs.~(\ref{q:xmed}) and (\ref{q:fu}). 
In particular, in the work of \citet{Coc12a}, many rates come from theory (TALYS code) \citep{TALYS} and have not been re-evaluated since then. 
The TALYS code is obviously not well adapted to this low mass region where level densities are too low to justify the Hauser-Feshbach
approach. However, due to the lack of experimental data it can be used as a first guess for the hundred of reaction rates that are needed.
To estimate the uncertainty associated with TALYS rates in the relevant ranges of masses and temperature, \citet{Coc12a} compared them
with experimentally determined reaction rates \cite{NACRE,Eval2} and found that the differences do not exceed three orders of magnitude
{\em at BBN temperatures} (see Figs.~ 1--11 and 16--21 in \citet{Coc12a}).
In their sensitivity study, they were not found to influence significantly the results, when {\em individually} multiplied by factors up to 
$10^{\pm3}$, except for ten reactions whose rates were re-evaluated.  
Hence, for the rates, labeled "TALYS" in Table~4 of \citet{Coc12a},  we use here uncertainty factors of $f$=100 (i.e. $\sigma=\ln(100)$). 
(Those,  labeled "TALYS" in bold face in the same Table were re-evaluated so that calculated uncertainties are available.)
For rates provided without calculated uncertainty, labeled e.g. "CF88", "MF89", "Wag69",\ldots, (see Refs. in \citet{Coc12a})
we generally adopted $f$=3, except when the uncertainty is not provided but is obviously smaller e.g. "Ham10" \cite{Ham10} or "Nag06" where
we adopted a 40\% uncertainty ($f$=1.4). 
This may look arbitrary, but one of the main goal of this work is to identify potentially
influential reaction rates that may have to be improved in a subsequent stage.

\section{Results concerning \hli}

\begin{figure}
\begin{center}
\vskip -3.3cm
 \includegraphics[width=.8\textwidth]{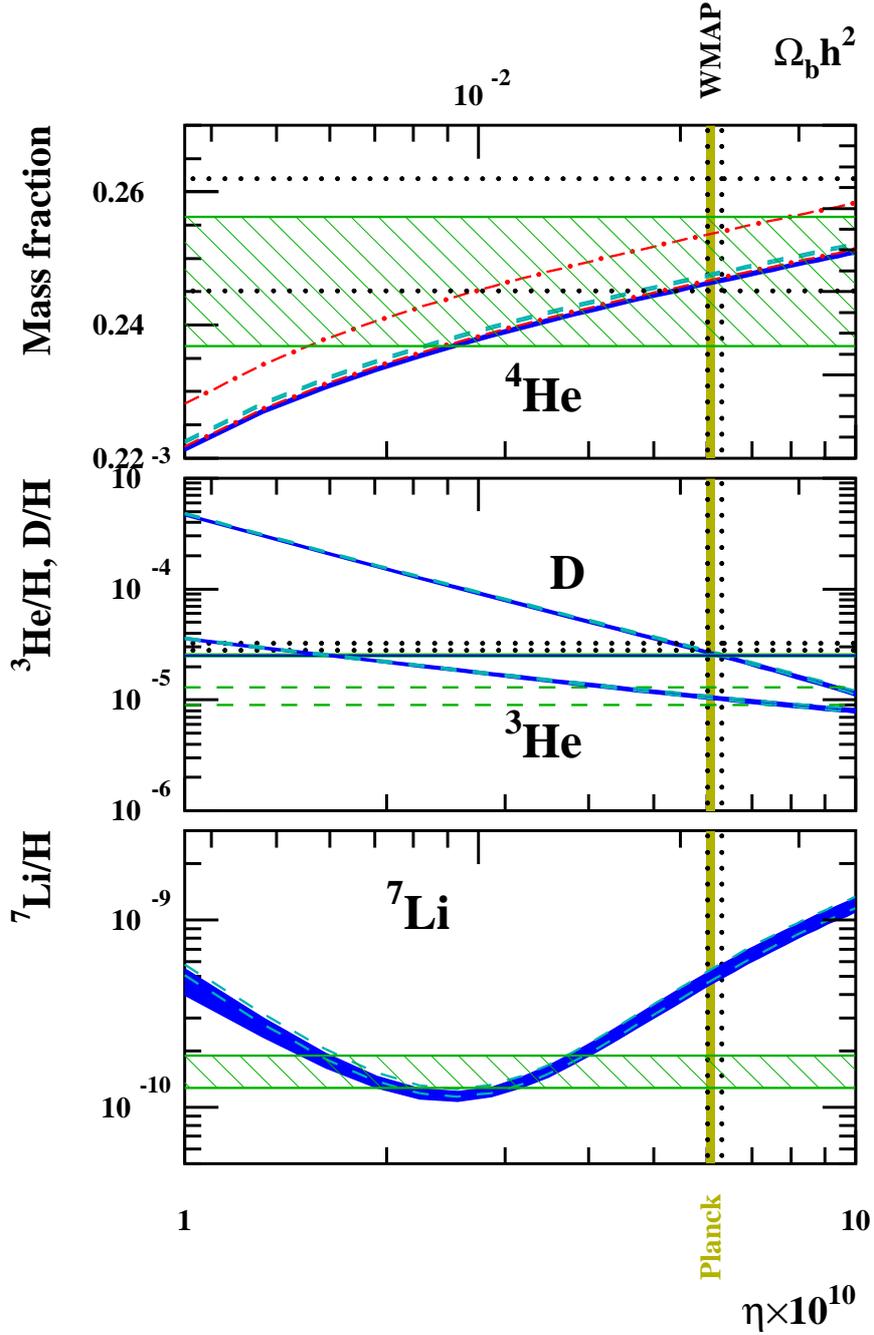}
\caption{\label{f:heli} (Color online)  \hli\ abundances as a function of $\eta$ calculated by Monte Carlo with the updated full network (dark blue) or 
with the reduced network as in Ref. \cite{CV10} (light blue dashed). The vertical areas correspond i)  to the WMAP (dot, black) and ii) {\it{Planck}} (solid, yellow) 
baryonic densities.
The horizontal areas (hatched green) represent the adopted observational abundances while  
the horizontal dotted lines correspond to those previously used \cite{aver12,olive2012}. 
The (red) dash--dotted lines correspond to $Y_p$ calculated with $N_{\rm eff}= 3.30 \pm0.27$ derived
from the CMB  \citep{Planck13}.  
}
\end{center}
\end{figure}

Figure~\ref{f:heli} displays the \hli\ abundances calculated as a function of $\eta$ by Monte Carlo with the full network,
and evaluated rate uncertainties following \citet{Coc12a}, compared to our previous work with a reduced network \citep{CV10}.
At $\eta_\mathrm{CMB}$, the differences are hardly visible except for \qua, due to the updated neutron lifetime.
Comparison between columns 2 and 5 in Table~\ref{t:hli} shows the evolution of the yields from \citet{CV10} with 
the first WMAP results \citep{Spe07} to  the recent {\it{Planck}} data \citep{Planck13}. The reduced uncertainty on D/H is a direct consequence
of the reduced uncertainty on \obh\ while \sep\ uncertainty is still dominated by nuclear uncertainty on the
$^3$He($\alpha,\gamma)^7$Be rate.

Figure~\ref{f:heli} displays the abundances as a function of $\eta$ and Table~\ref{t:obs} those at the {\it{Planck}} baryonic density, both for $N_{\rm eff}= 3$,
as  defined by Eq. (5.5).  
When using the last evaluation of $Y_p$ \citep{aver13} deduced from observations, we obtain  $2.19\leq N_{\rm eff} \leq 3.63$ at {\it{Planck}} baryonic density.
This interval is given after corrections to the weak rates \cite{Dic82,Lop99,Man05} have been introduced as discussed in \S~\ref{s:tables}.  
It includes the correction for non--instantaneous neutrino decoupling in the presence of oscillations, introduced as  a constant shift
$\Delta Y_p=+0.0002$ \citep{Man05} instead of a slight increase of  $N_{\rm eff}= 3$ since, if this approximation works for \qua,  the change for the other
nuclides is exactly in the opposite direction of the true one (see \citet{Man05} for details). 
In Figure~\ref{f:heli} we also display for visual inspection the results obtained for the limits on effective number of  neutrino family  $N_{\rm eff}= 3.30 \pm0.27$ derived
from the CMB only confidence interval \citep{Planck13}.  

Finally in Table~\ref{t:obs}, a comparison between this work and the last observational data is proposed; an overall consistency between standard BBN calculation and the observational constraints is presented except for lithium, as explained above: the discrepancy remains of the order of 3.  

\begin{table}[htbp!] 
\caption{Comparison with observations}
\begin{center}
\begin{tabular}{ccc}
\hline
   & This work & Observations   \\
\hline
$Y_p$*   & 0.2461--0.2466 & $ 0.2465 \pm 0.0097$ \\
\deu/H   ($ \times10^{-5})$&  2.57--2.72 & $2.53 \pm 0.04$\\
\tro/H    ($ \times10^{-5}$) &  1.02--1.08   &  $1.1 \pm 0.2 $  \\
\sep/H ($\times10^{-10}$)  &   4.56--5.34 & $   1.58  ^{+0.35}_{-0.28}$ \\
\hline
\end{tabular}\\
\; * See text in \S~\ref{s:tables}.\end{center}
\label{t:obs}
\end{table}

\section{Results concerning \six, \neu, B, C}
\label{s:cno}

\begin{figure}[htb!]
\begin{center}
 \includegraphics[width=.8\textwidth]{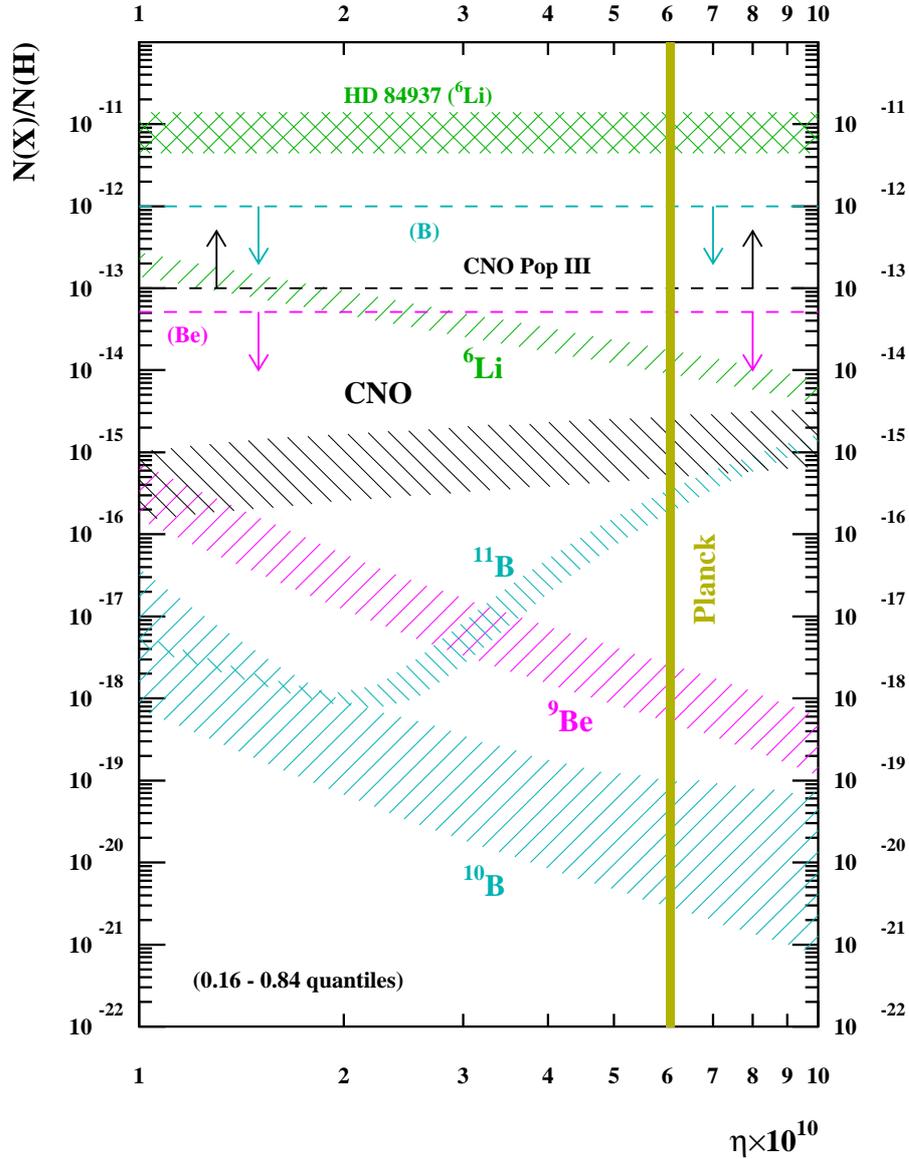}
\caption{ Standard big bang Nucleosynthesis predictions for the abundances of  \six, Be, B and CNO isotopes as a function of baryonic density,
compared to some observations.
The horizontal lines and areas correspond to the \six\ observation in HD 84937 (green), beryllium (magenta) and boron (light blue) upper 
limits, and CNO lower limit to affect Population III stars; see text. }
\label{f:bebc}
\end{center}
\end{figure}

Figure~\ref{f:bebc} displays the \six, \neu, \dix, \onz\ and CNO abundances  calculated as a function of $\eta$ 
including our estimated uncertainties from the 
Monte Carlo, and a comparison with observations. The displayed uncertainties are obtained by calculating 
for each  value of $\eta$, the 0.16 and  0.84 quantile \citep{Eval1} of the distributions.  The corresponding
confidence intervals at $\eta_\mathrm{CMB}$ are displayed in Table~\ref{t:lic} and are orders of magnitude
below observations (\S~\ref{s:obslibeb}).

\begin{table}[h!]
\caption{\label{t:lic}  \six\ to CNO primordial abundances by number.}
\begin{center}
\begin{tabular}{ccc}
\hline
  &  Ref. \citep{Coc12a} &  This work \\
\hline
\six/H ($\times10^{-14}$) & 1.23  & 0.90--1.77 \\
\neu/H ($\times10^{-19}$)  &  9.60 &    5.10--26.3   \\
\onz/H  ($\times10^{-16}$)  &  3.05 &    1.85--3.56  \\
CNO/H  ($\times10^{-16}$) &  7.43 & 4.94--28.5  \\
\hline
\end{tabular}\\
\end{center}
\end{table}

\begin{figure}[htb!]
\begin{center}
 \includegraphics[width=.8\textwidth]{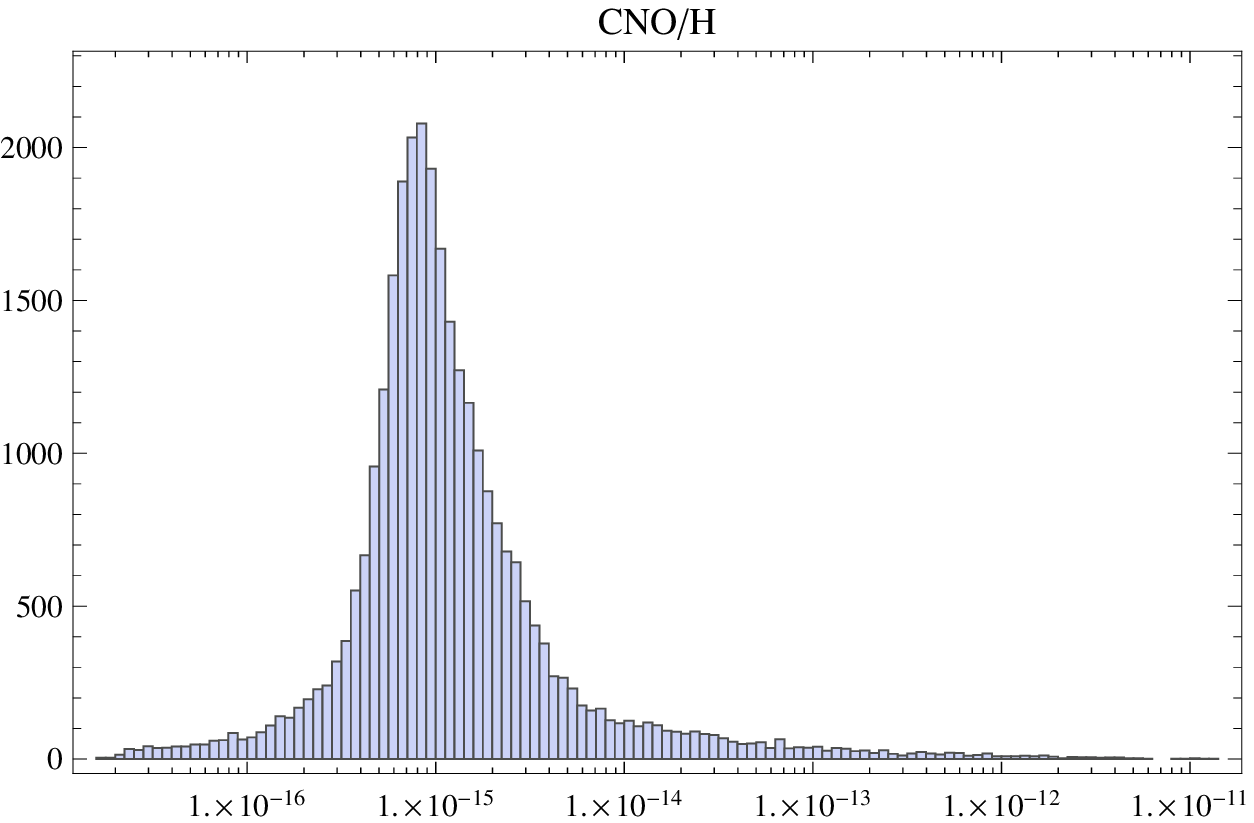}
\caption{\label{f:hcno} CNO/H distribution showing that high values are obtained in a non negligible proportion.}
\end{center}
\end{figure}

Figure~\ref{f:hcno} displays the histogram of CNO/H obtained from our Monte Carlo calculation, from which it is possible to extract
the  0.16, 0.5 and  0.84 quantile, respectively given by 4.94$\times10^{-16}$, 9.63$\times10^{-16}$ and 2.85$\times10^{-15}$. 
This is very close to the range CNO/H = $(0.5-3.)\times10^{-15}$ estimated by \citet{omeg11} from the results of \citet{Coc12a}.
However, at high value, the tail of the distribution ($\approx$3\%) extends to values much above  the  CNO/H = $10^{-13}$ limit.
At first, it seems straightforward to extract the subset of events, for which  e.g. CNO/H$>10^{-13}$,  and
examine the corresponding sampled reaction rates (i.e. the $p_k$'s in Eq.~\ref{q:ln}) that are stored together with the final abundances in a database. 
However, since all $\approx$400 $p_k$'s  are different from one realisation to the other, it was not possible to identify combination of reaction rates 
that produced such an effect.  To identify those combinations of reaction rates that allows such high value, we relied upon the calculated correlations 
between rates and yields.

This method is complementary to the one used by \citet{Coc12a}, in which a single reaction was tested at a time  by changing its rate
by factors of 10$^n$, $n= -3,-2,-1, 1, 2, 3$. Here {\em all rates are simultaneously changed by factors, different for each reaction}, and randomly
sampled as described above. This allows to identify sub-networks rather than individual reactions and takes into account the different
uncertainty factors.   
Results are displayed in Tables~\ref{t:cqua}--\ref{t:ccno} when their absolute value exceed 10\%. \footnote{
We use this simple criterium but are aware that more sophisticated criteria need to be developed: see e.g. Fig.~6 in \citet{Stats}.}
Note however that a higher sensitivity, as calculated in \citet{CV10,Coc12a}, does not necessarily correspond to a 
higher correlation as calculated here. Our previous sensitivity studies assumed an arbitrary  $\pm$15\% rate variation \citep{CV10}
or a factor of up to 1000 rate variation \citep{Coc12a}. Here, while, when sampling the rates, we still allow for large arbitrary rate variations 
for reactions with no documented rate uncertainties, we restrict the variations to evaluated rates and associated uncertainties when available. 
These latter reactions include e.g. those evaluated by \citet{Des04} or those identified as influential in a first step, but evaluated in a second
step by  \citet{Coc12a}.
For instance, the most influential rate on $^7$Li+$^7$Be is  $^1$H(n,$\gamma)^2$H \citep{CV10} but its rate uncertainty is very small \citep{And06} 
so that it does not appear in the table, contrary to the next most influential,  $^3$He($\alpha$,$\gamma$)$^7$Be whose rate uncertainty is still not negligible 
and thus  appears in Table~\ref{t:csep}.

The reactions that appear most correlated with the isotopes lighter than C (Tables~\ref{t:cqua}--\ref{t:conz}) are among those found in previous studies. 
In Table~\ref{t:ccno}, it appears that  besides the already known influential reactions on CNO production \citep{Coc12a} 
[$^7$Li(d,n)2$^4$He,  $^{12}$B(p,$\alpha$)$^9$Be, $^8$Li($\alpha$,n)$^{11}$B, 
$^{13}$C(d,$\alpha$)$^{11}$B],  
new influential reactions are found. It indicates a new possible path for  CNO production
involving $^{10}$Be, namely:  
$^{10}$Be(p,$\alpha$)$^7$Li, $^{10}$Be($\alpha$,n)$^{13}$C,   
$^7$Li(t,$\gamma$)$^{10}$Be,   $^8$Li(t,n)$^{10}$Be,   $^{10}$Be(t,n)$^{12}$B and 
$^{10}$Be(p,t)2$^4$He.
Note that all these new reactions involve radioactive isotope(s) [$^3$H, $^8$Li and $^{10}$Be] in the initial state, hence the absence of direct experimental data.
From this analysis, it is obvious that a new chain of reactions leading to CNO via $^{10}$Be need further attention as it could, depending on the cross-sections, 
provide a more efficient  source of CNO.

\twocolumn

\begin{table}[htbp!] 
\caption{\label{t:cqua} Correlations with \qua}
\begin{center}
\begin{tabular}{cc}
\hline
 Reaction & $C_{{\rm{He4}},k}$   \\
\hline
\hline

\hline
    1/$\tau_{\rm{n}}$          &     -0.9677\\
    $^3$He(t,np)$^4$He    &        0.1151\\
   D(d,n)$^3$He          &    0.1282\\
   D(d,p)$^3$H           &    0.1296\\
\hline
\hline
\end{tabular}\\
\end{center}
\end{table}

\begin{table}[htbp!] 
\caption{\label{t:cdeu} Correlations with \deu}
\begin{center}
\begin{tabular}{cc}
\hline
 Reaction & $C_{{\rm{D}},k}$   \\
\hline
\hline
\hline
  D(p,$\gamma$)$^3$He         &    -0.7790\\
   D(d,n)$^3$He        &     -0.4656\\
   D(d,p)$^3$H           &   -0.4082\\
\hline
\hline
\end{tabular}\\
\end{center}
\end{table}

\begin{table}[htbp!] 
\caption{\label{t:ctro} Correlations with \tro}
\begin{center}
\begin{tabular}{cc}
\hline
 Reaction & $C_{{\rm{He3}},k}$   \\
\hline
\hline
 \hline
  D(p,$\gamma$)$^3$He        &      0.6699\\
   D(d,n)$^3$He         &     0.1640\\
   D(d,p)$^3$H          &    -0.1897\\
   $^3$He(d,p)$^4$He      &      -0.6841\\
\hline
\hline
\end{tabular}\\
\end{center}
\end{table}

\begin{table}[htbp!] 
\caption{\label{t:csep} Correlations with \sep}
\begin{center}
\begin{tabular}{cc}
\hline
 Reaction & $C_{{\rm{Li7}},k}$   \\
\hline
\hline
\hline
   $^7$Be(n,$\alpha$)$^4$He      &      -0.3057\\
   $^7$Be(d,p)2$^4$He     &      -0.2079\\
  D(p,$\gamma$)$^3$He         &     0.4043\\
  D(d,n)$^3$He          &    0.1547\\
  $^3$He(d,p)$^4$He      &      -0.2232\\
  $^3$He($\alpha$,$\gamma$)$^7$Be       &      0.7107\\
\hline
\hline
\end{tabular}\\
\end{center}
\end{table}

\begin{table}[htbp!] 
\caption{\label{t:csix} Correlations with \six}
\begin{center}
\begin{tabular}{cc}
\hline
 Reaction & $C_{{\rm{Li6}},k}$   \\
\hline
\hline
\hline
 D($\alpha$,$\gamma$)$^6$Li        &      0.9835\\
  $^6$Li(p,$\alpha$)$^3$He      &      -0.1333\\
\hline
\hline
\end{tabular}\\
\end{center}
\end{table}

\begin{table}[htbp!] 
\caption{\label{t:cneu} Correlations with \neu}
\begin{center}
\begin{tabular}{cc}
\hline
 Reaction & $C_{{\rm{Be9}},k}$   \\
\hline
\hline
\hline
 $^7$Li($^3$He,p)$^9$Be    &       0.2820\\
  $^7$Li(t,n)$^9$Be         &    0.8910\\
\hline
\hline
\end{tabular}\\
\end{center}
\end{table}

\begin{table}[htbp!] 
\caption{\label{t:conz} Correlations with \onz}
\begin{center}
\begin{tabular}{cc}
\hline
 Reaction & $C_{{\rm{B11}},k}$   \\
\hline
\hline
\multicolumn{2}{c}{\onz+$^{11}$C}\\
\hline
  D(p,$\gamma$)$^3$He      &        0.1081\\
  $^3$He($\alpha$,$\gamma$)$^7$Be       &      0.1621\\
 $^7$Be($\alpha$,$\gamma$)$^{11}$C       &      0.2188\\
 $^{11}$C(n,$\alpha$)2$^4$He     &      -0.9069\\
\hline
\hline
\end{tabular}\\
\end{center}
\end{table}

\begin{table}[htbp!] 
\caption{\label{t:ccno} Correlations with CNO}
\begin{center}
\begin{tabular}{cc}
\hline
 Reaction & $C_{{\rm{CNO}},k}$   \\
\hline
\hline
\multicolumn{2}{c}{CNO}\\
\hline
 $^{12}$B(p,$\alpha$)$^9$Be       &     -0.1030\\
  $^{10}$Be(p,$\alpha$)$^7$Li       &    -0.2479\\
 $^{10}$Be($\alpha$,n)$^{13}$C     &       0.2175\\
 $^7$Li(t,$\gamma$)$^{10}$Be       &     0.1363\\
 $^8$Li($\alpha$,n)$^{11}$B        &     0.1449\\
 $^8$Li(t,n)$^{10}$Be       &     0.1958\\
 $^{10}$Be(t,n)$^{12}$B     &       0.1139\\
 $^{13}$C(d,$\alpha$)$^{11}$B       &     -0.2378\\
 $^{10}$Be(p,t)2$^4$He    &      -0.1297\\
 $^7$Li(d,n)2$^4$He     &      -0.1806\\
\hline
\hline
\end{tabular}\\
\end{center}
\end{table}

\onecolumn

\begin{figure}[htb!]
\begin{center}
 \includegraphics[width=.8\textwidth]{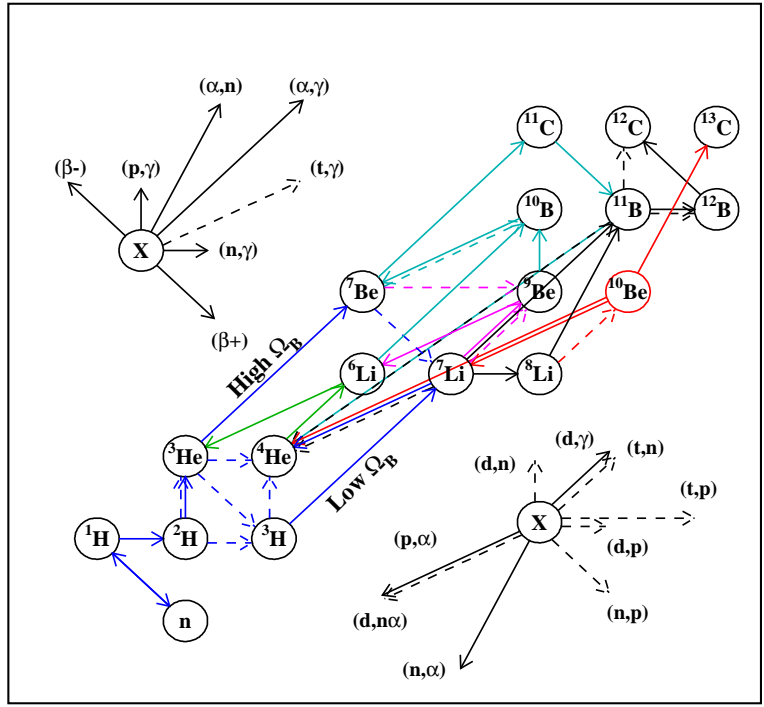}
\caption{\label{f:netw} (Color online) BBN nuclear network with in red the new possible paths to CNO.
}
\end{center}
\end{figure}

These reactions were not identified  in our previous work \cite{Coc12a}, because we varied the rates, {\em one at a time}, 
(by factors of $10^n$ with $n$ varying from $-3$ to 3 by steps of one unit). 
For instance,  when increasing the $^8$Li(t,n)$^{10}$Be {\em or} $^{10}$Be($\alpha$,n)$^{13}$C rates by a factor of 1000,   
the CNO abundance only increase by 30\% \cite{Coc12a} while if {\em both} rates are increased by the same factor, CNO/H is found
to be higher by a factor of 200. This is the purpose of the following analysis to identify such potential new paths. 
Hence, we first allowed {\em all rates to vary simultaneously and independently} according to lognormal distributions.
Now, to better identify the chains of reactions that may lead to an increased CNO production, we allow the rates of the 6 newly identified reactions listed above 
to vary  within a few orders of magnitude as in Ref. \cite{Coc12a} but considering {\em all} possible combinations. 
We chose factors of 10$^{\pm2}$ variations on rates w.r.t. TALYS calculated rates which are consistent with our comparison between TALYS and experimentally measured
rates \citep{Coc12a} and select those combinations of factors that leads to a CNO/H production higher than 10$^{-13}$, the minimum value to affect Pop III stars \cite{Eks08}.  
In Table~\ref{t:comb} are displayed the  9 combinations (out of $3^6=729$ combinations ) of signs  in the exponent of the10$^{\pm2}$ factors (with "0" meaning no rate variation) for
which CNO/H $>10^{-13}$. (Obviously, we could have reduced the number of combinations: taking higher/lower rates for production/destruction of $^{10}$Be but
we found it was not worth the trouble.)
Table~\ref{t:comb}, shows that the $^7$Li(t,$\gamma$)$^{10}$Be and $^{10}$Be(t,n)$^{12}$B reactions are not essential since whatever the exponent (-2, 0 or +2) of the 
variation factor, the result is not significantly affected.
On the contrary, the combination of higher rates for $^{10}$Be($\alpha$,n)$^{13}$C  and $^8$Li(t,n)$^{10}$Be together with lower rates for $^{10}$Be(p,$\alpha$)$^7$Li
and $^{10}$Be(p,t)2$^4$He result in a substantial increase in primordial CNO production. From this combinatorial analysis we could, hence,
separate within the set of 6 reaction rates that were weakly correlated to CNO/H those 4 that really matter.  
The factors of 10$^{\pm2}$ variation w.r.t. TALYS rates is conservative (see \S~\ref{s:rates}) so that even higher CNO yields can be expected.
Experimental investigations of these four reactions (Fig.~\ref{f:netw}) are hence highly recommended.

Afterwards, the production of this short list of reactions involving $^{10}$Be for the production of CNO may seems straightforward but  
dealing with a relatively large and dense network (i.e. including p, n, $\alpha$, d, t and \tro\ induced reactions) is not so easy.
First, among all the reaction paths that could connect A$<8$ to CNO nuclei, the study of correlations allowed us first to identify,
the six reactions of Table~\ref{t:comb} that did not show up in our (or any other) simpler sensitivity analysis \cite{Coc12a}
of the same network. Second, the extensive combinatorial analysis allowed us to finally select out four reactions.

\begin{table}[htbp!] 
\caption{\label{t:comb} Each column correspond to a combination of multiplicative factors ("-" for 10$^{-2}$, "0" for 10$^0$ and "+" for 10$^{+2}$) which applied 
simultaneously to all the six TALYS reaction rates lead to CNO/H$>10^{-13}$.}
\begin{center}
\begin{tabular}{cccccccccc}
\hline
 Reaction & \multicolumn{9}{c}{CNO/H$>10^{-13}$} \\
\hline
 \hline
  $^{10}$Be(p,$\alpha$)$^7$Li &-&-&-&-&-&-&-&-&-\\
 $^{10}$Be($\alpha$,n)$^{13}$C     &+&+&+&+&+&+&+&+&+\\
 $^7$Li(t,$\gamma$)$^{10}$Be        &-&-&-&0&0&0&+&+&+\\
 $^8$Li(t,n)$^{10}$Be      &+&+&+&+&+&+&+&+&+\\
 $^{10}$Be(t,n)$^{12}$B      &-&0&+&-&0&+&-&0&+\\
  $^{10}$Be(p,t)2$^4$He     &-&-&-&-&-&-&-&-&-\\
\hline
\hline
\end{tabular}\\
\end{center}
\end{table}

This work has updated the BBN predictions in order to take into account the most recent developments concerning the cosmological framework (i.e. the cosmological parameters determined from the recent CMB {\it{Planck}} experiment). It demonstrates that these predictions are robust for the lightest elements. It shows also that the modification of these parameters in the range allowed cannot alleviate the lithium problem; concerning primordial CNO production we show that higher CNO yields can be expected: the four reaction rates  $^{10}$Be($\alpha$,n)$^{13}$C, $^8$Li(t,n)$^{10}$Be,  $^{10}$Be(p,$\alpha$)$^7$Li
and $^{10}$Be(p,t)2$^4$He,  could be investigated to test this result.

Finally, we want to emphasize the use of statistical methods in BBN  have lead to the identification of a possible new path to CNO. 
For this, we have used the simple Pearson's correlation coefficient to discriminate important reactions.  This is obviously a first step: more
elaborate statistical techniques could be developed and also applied to other nucleosynthesis sites \citep{Stats}.

\acknowledgments
This work made in the ILP LABEX (under reference ANR-10-LABX-63) was supported by French state funds managed by the ANR 
within the Investissements d'Avenir programme under reference ANR-11-IDEX-0004-02 and by the ANR VACOUL, ANR-10-BLAN-0510. 
We are indebted to Christian Iliadis and Richard Longland for many fruitful discussions, and in  particular on statistical issues concerning 
reaction rates and nucleosynthesis.
We thank warmly the referee for his remarks and very useful comments.

\appendix

\section{Appendix: Relation between $\eta$ and \obh}

Here, for precise comparison with other works that quote $\eta$ numerical values rather than \obh\ ones, we recall here the numerical
relationship between the two. This was calculated previously by \citet{Ste06}; here we detail our own calculation.

What is important for BBN is the baryonic density $\Omega_{\mathrm{b}}{\cdot}\rho_{0,C}$ where $\rho_{0,C}$ is the present day critical density given by
 (numerical values of physical constants are taken from \citet{PDG12}, atomic masses from \citet{Ame12}):
\begin{equation}
\rho_{0,C}=\frac{3H_0^2}{8{\pi}G}=1.87847\times10^{-29}\;h^2\;
\mathrm{g/cm^3},
\label{q:critic}
\end{equation}
that allows for the calculation of 
\begin{equation}
\rho_{\mathrm b}(t)=\left[1.87847\times10^{-29}\;\mathrm{g/cm^3}\right]\;\times\;\Omega_{\mathrm{b}}h^2a^{-3}(t)\;,
\label{q:critik}
\end{equation}
to be used in the network calculations.
The photon density (number/cm$^3$; $T_0$ = 2.7255~K; $\zeta(3)$ = 1.20206) is:
\begin{equation}
n_{\gamma}(T)=\frac{2\zeta(3)}{\pi^2}\left(\frac{\mathrm{k_B}T}{\hbar c}\right)^3=410.73\left(\frac{T}{T_0}\right)^3\;
\mathrm{cm^{-3}}
\end{equation}
The number of baryon per photon is thus given by:
\begin{equation}
\eta=\frac{\rho_{0,\mathrm{b}}}{n_{\gamma}(T_0)\bar{M}}\equiv\frac{3H_0^2}{8{\pi}G}\frac{\pi^2}{\zeta(3)}\left(\frac{\hbar c}{\mathrm{k_B}T_0}\right)^3
\left(M_p(1-Y_p)+\frac{M_\alpha}{4}Y_p\right)^{-1}\Omega_\mathrm{b}
\end{equation}
where $\bar{M}$ is the mean baryon mass
\begin{equation}
\bar{M}=M_p(1-Y_p)+\frac{M_\alpha}{4}Y_p=(1.6735 -0.0119\;Y_p)\times10^{-24}\;\;(g).
\end{equation}
So that the relation between $\eta$ and \obh\ are:
\begin{equation}
\eta=2.7381\times10^{-8}\;\Omega_{\mathrm{b}}h^2\;\;\mathrm{or}\;\;\Omega_{\mathrm{b}}h^2=3.6521\times10^{7}\;\eta
\end{equation}
for $Y_p$=0.27 (solar, $\bar{M}=1.6703\times10^{-24}$ g) or
\begin{equation}
\eta=2.7377\times10^{-8}\;\Omega_{\mathrm{b}}h^2\;\;\mathrm{or}\;\;\Omega_{\mathrm{b}}h^2=3.6528\times10^{7}\;\eta
\end{equation}
for $Y_p$=0.246 (BBN, $\bar{M}=1.6706\times10^{-24}$ g).
When using the same values for $T_0$  (2.725 instead of 2.7255~K here) and  $Y_p$ (0.27), our result differs by less than 0.004\% 
from the \citet{Ste06} one. The uncertainty on the present day CMB temperature ($T_0= 2.7255\pm0.0006$ \citet{PDG12}) induces an uncertainty
of less than 0.07\% on the calculated coefficient.


\end{document}